# Hybrid subterahertz atmospheric pressure plasmatron for plasma chemical applications

*Sintsov S.V.[1,2,a], Vodopyanov A.V.[1,2], Mansfeld D.A.[1], Fokin A.P.[1], Ananichev A.A.[1], Goryunov A.A.[2], Preobrazhensky E.I.[1], Chekmarev N.V.[1], Glyavin M.Yu.[1].*

1) A.V. Gaponov-Grekhov Institute of Applied Physics of the Russian Academy of Sciences (IAP RAS), Nizhny Novgorod 603950, Russia
2) Lobachevsky State University of Nizhny Novgorod (UNN), Nizhny Novgorod 603022, Russia

a) Author to whom correspondence should be addressed: sins@ipfran.ru

## Abstract

This paper presents the results of an experimental study of a new hybrid plasmatron scheme, which was used to realize a gas discharge at atmospheric pressure supported by continuous focused submillimeter radiation with a frequency of 263 GHz. The implemented design allowed organizing a self-consistent interaction between submillimeter radiation and the supercritical plasma in a localized area both in terms of gas flow and electrodynamic. It is experimentally shown that the gas discharge absorbs up to 80% of the introduced submillimeter radiation power. The hybrid subterahertz plasmatron as an effective reactor for non-equilibrium plasma chemical processes was tested for the atmospheric nitrogen fixation.

## Introduction

One of the least investigated areas in the field of microwave discharge physics is the study of the possibility to create nonequilibrium plasma at high, close to atmospheric, pressure using electromagnetic radiation of millimeter and submillimeter wavelength range [1-7]. Conducting such studies has become possible relatively recently and is associated with the development of powerful and reliable sources of electromagnetic radiation - gyrotrons [8-14]. Recent advances in the development of electrovacuum microwave generators open up a wide range of possibilities for exploring new applications of powerful sub-terahertz and terahertz radiation [9-14].

In theory, the use of high-power electromagnetic radiation of submillimeter wavelength range allows to obtain at high pressure inhomogeneous collisionless plasma ($\nu_{em} \ll \omega$) with peak electron density exceeding the cut-off value for the field frequency ω [15-17]. In the local region where the plasma dielectric permittivity is close to 0, the electric field strength of the wave can be significantly increased due to nonlinear resonance electrodynamic effects, which in turn causes the formation of a plasma with a significantly non-equilibrium distribution of temperature characteristics [18-25].

One of the most widely used approaches to the design of high-pressure microwave plasmatrons is the use of electrodynamic waveguide structures, with characteristic dimensions on the order of the wavelength of electromagnetic radiation [25-29]. The gas discharge localized by the walls of the plasmatron's electrodynamic structure is supported by microwave radiation in the flow of plasma-forming gas, which is also introduced into the waveguide structure. Transition to submillimeter wavelengths allows to minimize the size of the plasmatron's electrodynamic structures, which makes it possible to increase the energy flux density in the area of discharge at a fixed power. On the other hand, this approach to the design of gas-discharge plasma sources supported by electromagnetic radiation of the subterahertz and terahertz ranges cannot be fully used due to the specific features of the high-pressure plasmatron's temperature regimes and the operating conditions of submillimeter radiation sources, whose operation may be disrupted in the presence of a reflected wave [30, 31]. For these reasons, quasi-optical electrodynamic systems are used to organize gas discharges supported by submillimeter radiation [1, 4, 5, 23]. The subterahertz beam is focused on a gas nozzle and the discharge is initiated in the gas flow propagating towards the wave vector. A similar scheme was realized in previous works on the study of an atmospheric pressure plasma torch supported in an opposing flow of a mixture of argon and carbon dioxide by continuous focused radiation with a frequency of 263 GHz [24, 32]. The main problem limiting the use of this type of discharge for plasma chemical applications is the significant reflection of electromagnetic radiation from the boundary of the plasma torch, which has a supercritical electron density. The discharge absorbed no more than 15% of the input power, which provided the creation of a plasma with a temperature of about 2500 - 3500 K. At the content of carbon dioxide in the plasma-forming mixture exceeding 5%, the discharge could not be sustained. Despite the essentially nonequilibrium character of carbon dioxide destruction realized in this type of discharge, the energy efficiency of the process, which characterizes the fraction of energy spent on dissociation of molecules, did not exceed 1% [32].

This paper presents the results of an experimental study of a new hybrid scheme for gas discharge organization at atmospheric pressure sustained by continuous focused submillimeter radiation with a frequency of 263 GHz. The key idea behind the design of the hybrid subterahertz plasmatron was to realize the joint interaction between submillimeter radiation and the subcritical plasma it sustains in a localized region, both in terms of gas flows and electrodynamics. In the new design, the plasma gas flow is co-directed with the wave vector of the electromagnetic field. The plasma and focused submillimeter radiation with a frequency of 263 GHz interaction was organized in a metallic electrodynamic structure having the shape of a truncated cone located in the region of the beam waist. As a result, absorption of submillimeter radiation by the discharge at the level of 70-80% was achieved. Hybrid plasmatron as an efficient reactor for plasma chemical processes was tested in the synthesis of nitrogen oxides ($NO$, $NO_2$) from atmospheric air.

**Experimental setup**

The main elements of the experimental setup are a gas-discharge chamber and a submillimeter radiation source. A gyrotron with an operating frequency of 263 GHz and power up to 1.1 kW in continuous wave mode was used as a source of electromagnetic radiation [13, 14]. Subterahertz radiation with linear polarization had a Gaussian intensity distribution in the beam cross section at the output. The radiation was transported and focused using a supersized quasi-optical transmission line. The first mirror directed the radiation into the chamber through the polyethylene input window. The hermetically sealed gas-discharge chamber was divided into two parts (Figure 1). In the upper part - a vacuum chamber in the form of a cube, there was a system of plasma-forming gases supply, an input window, a second parabolic mirror with an alignment system, a water flow calorimeter, and an observation window. After the second parabolic mirror, the submillimeter beam was focused into a spot whose position corresponds to the level of the lower face of the camera. The angular divergence of the beam was 60º, the cross-sectional diameter of the quasi-optical beam in the focal waist was 1.2 mm, the power density was up to 20 kW/cm$^2$, and the RMS value of the electric field was up to 2.7 kV/cm. The lower chamber had a second flow water calorimeter, a system of output and sampling of plasma-forming gases for their further component analysis. Between the upper and lower chambers there was a flange with a hermetically embedded electrodynamic structure, the walls of which formed a truncated cone. Both the plasma-forming gas and the electromagnetic radiation could get from the upper chamber to the lower chamber only through this electrodynamic horn. The angle at the top of the truncated conical structure is equal to the angular convergence of the beam, and the diameters of the bases were 5 and 32 mm. The smaller base transitioned into a thin cylindrical channel 22 mm long.

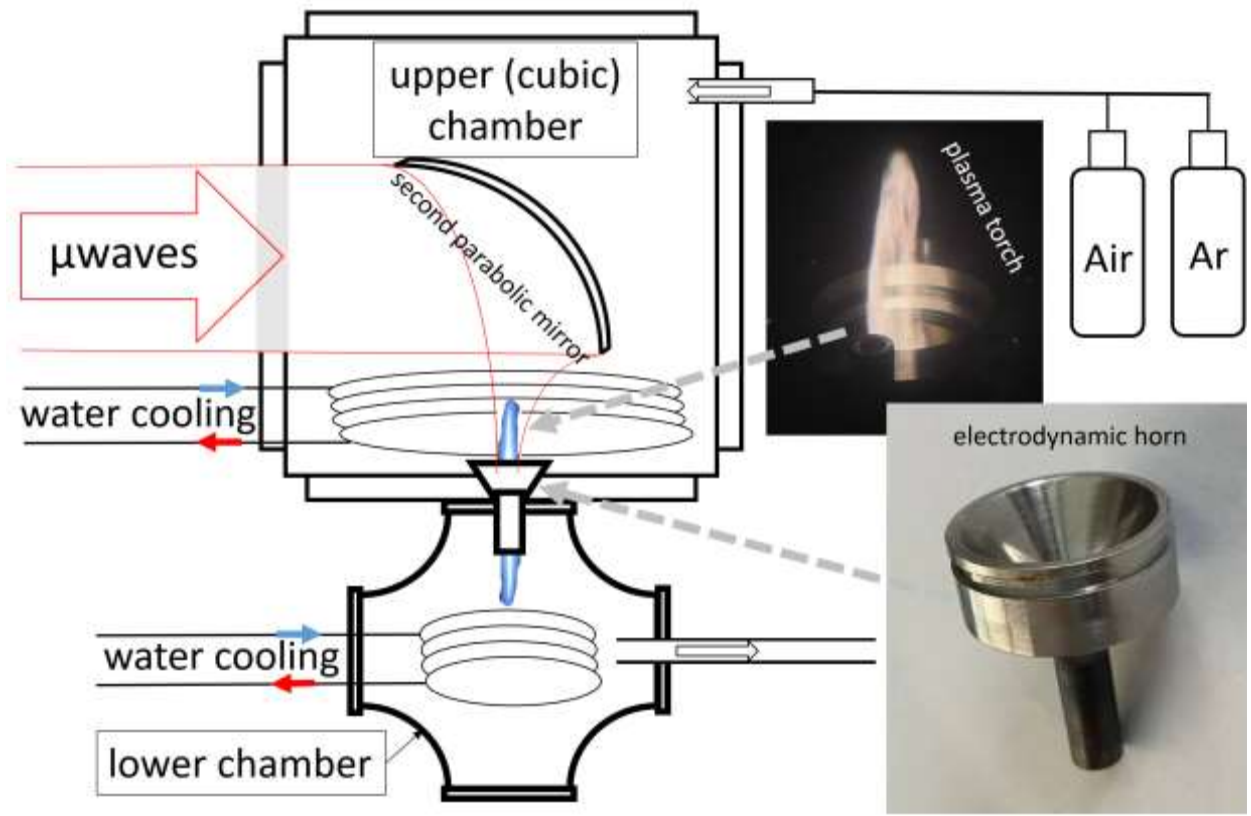

*Figure 1. Schematic diagram of the experimental setup.*

The focused submillimeter beam was coaxially introduced into the developed electrodynamic structure (Figure 2a). Precise alignment of the beam relative to the horn allowed to achieve moderate reflection of electromagnetic radiation in the upper chamber: in the absence of plasma, 90±2% of the input power of submillimeter radiation passed into the lower gas-discharge chamber. The upper hermetically sealed chamber was pumped with a plasma-forming gas mixture at atmospheric pressure. The injected gas flowed through the horn into the lower chamber. The plasma torch in the gas flow, co-directed with the wave vector of the focused submillimeter beam, was initiated inside the horn using a metal wire (Figure 2b). The size and shape of the discharge depended on the microwave heating power, composition and ratio of components of the plasma-forming mixture. In the argon flow, the plasma torch was initiated at a subterahertz radiation power of 190 W at both ends of the structure, i.e., it was elongated in both directions. The plasma torch extended toward the argon flow from the wide base of the horn consisted of filamentary plasma channels (Figure 3a). Its length was proportional to the input power and was up to 70 mm. We suppose that the length of the torch was determined by the level where the electric field achieved a value sufficient to sustain the plasma. The torch, elongated in the direction of the argon flow, came out of a thin cylindrical channel with an inner diameter of 5 mm, which is an extension of the narrow base of the truncated conical electrodynamic structure. Its length also depended on power and was up to 12 mm.

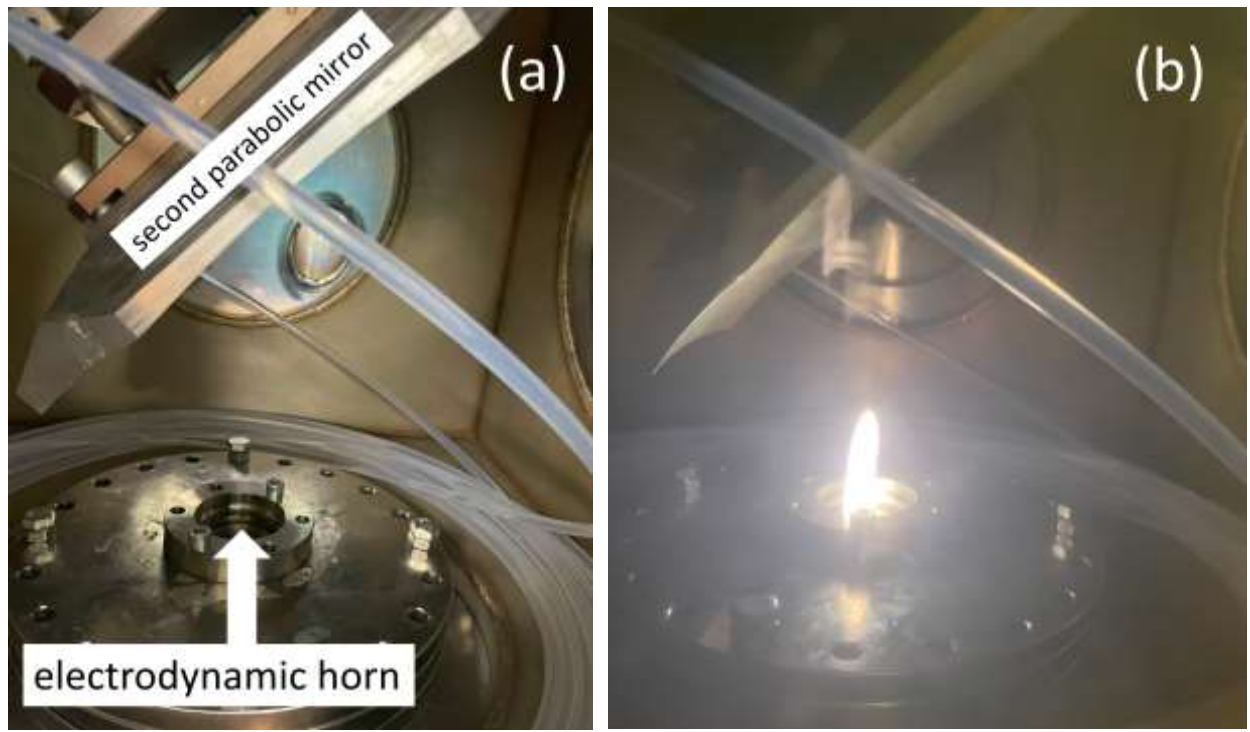

*Figure 2. Photographs of (a) the submillimeter focusing system and; (b) the plasma torch supported in an argon flow with focused submillimeter radiation.*

The addition of the molecular gas led to a decrease in the length of the plasma torch directed toward the gas flow and to a total retraction of the second torch (directed along the gas flow) into the cylindrical channel. Air, hydrogen and carbon dioxide were used separately as additives in the mixture with argon (Figure 3 b,c,d). Increasing the content of molecular gas in mixture with argon to 20-25% leads the discharge to the regime where both plasma torches are absent and the plasma is localized entirely inside the horn, because the highest electric field strength is achieved there. At a higher content of the molecular component in the mixture with argon, the gas discharge could not be initiated. In the experiments performed to sustain the discharge in the developed subterahertz plasmatron, the power absorbed by the plasma was measured. For this purpose, water calorimeters were placed in the upper and lower gas-discharge chambers. The calorimetric systems were made of fluoropolymer tubes and had separate water circuits. In each calorimetric circuit, the water flow rate and the temperature difference between the inlet and outlet were measured continuously, which made it possible to calculate the electromagnetic radiation power absorbed by the water with an accuracy of 20 W. It has been shown experimentally that the realized calorimetric system can register 94±2% of the input power of the subterhertz radiation. The unregistered part of the power was probably absorbed in the walls of the gas-discharge chamber or came out through the polyethylene window. The integral optical emission spectra of the plasma torch elongated toward the plasma-forming gas flow were registered through the quartz window located in the upper gas-discharge chamber. A two-channel spectrometer S150 Duo from SOL Instruments was used. The first channel, used to register the observation spectra in the 200 - 1100 nm range, has a diffraction grating with a period of 200 l/mm. The second channel has a diffraction echelle grating with a period of 75 l/mm. In the narrow tunable spectral region, the use of the echelle channel provided a resolution of 7 to 30 pm, depending on the analyzed range, so it was used to record the shapes of selected emission lines. Component analysis of gas samples was performed using an Agilent 6890/MSD 5973N chromatography-mass spectrometer. Based on the results of the component analysis, the efficiency of nitrogen fixation in atmospheric air in the realized subterahertz discharge was evaluated.

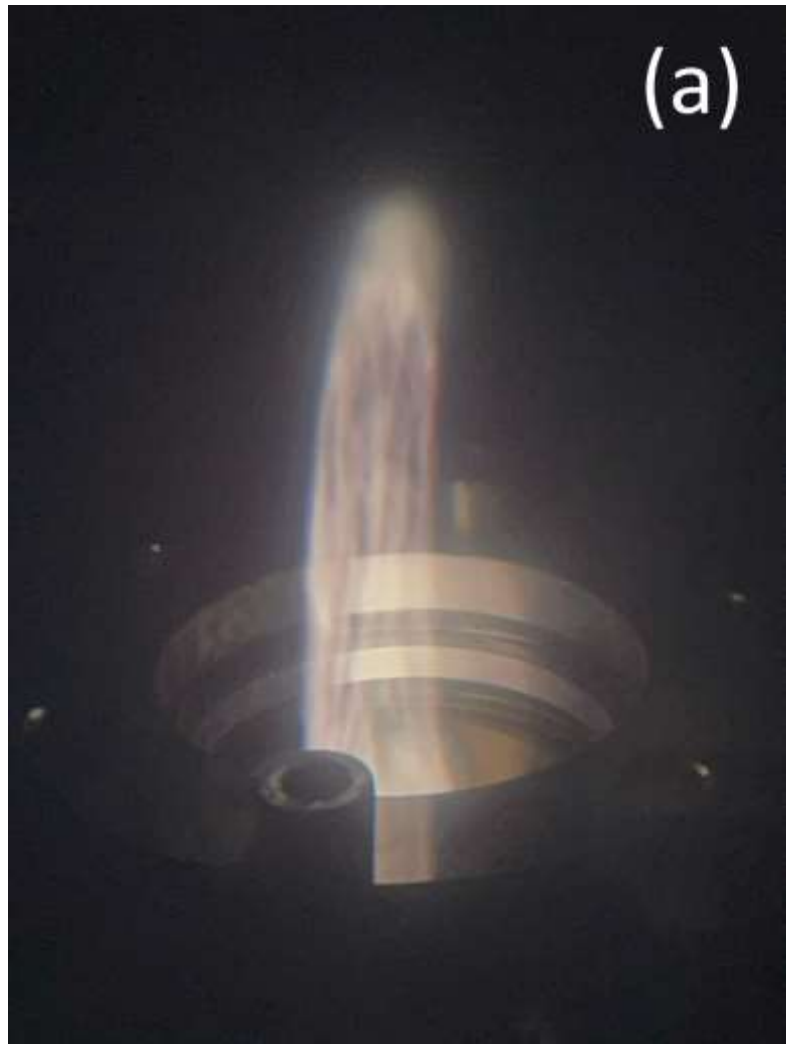
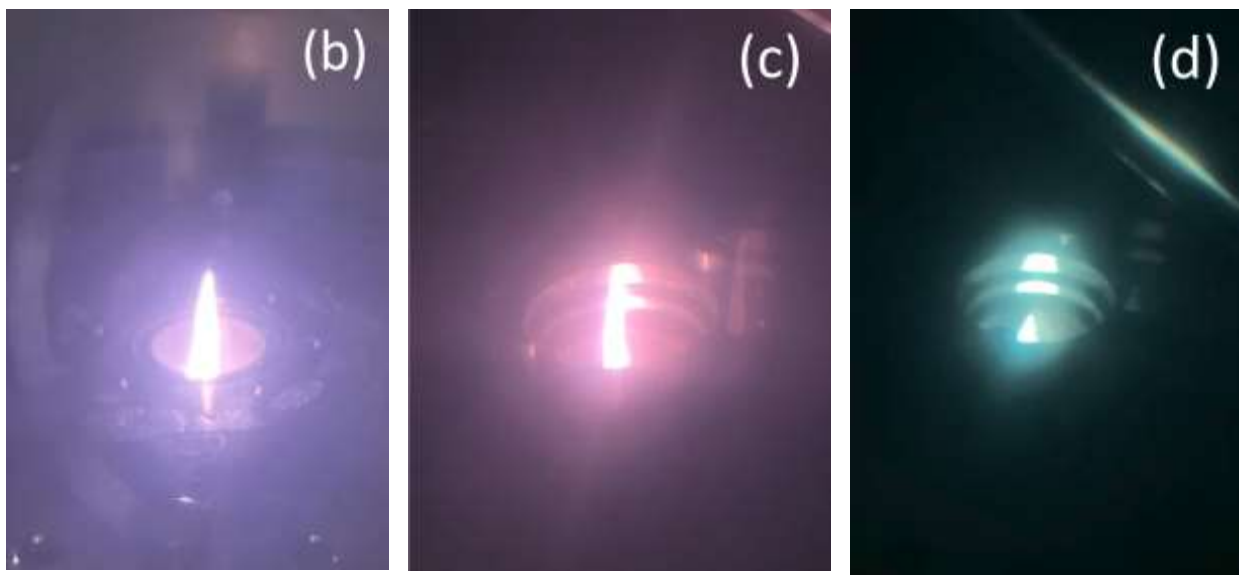

*Figure 3. Photographs of plasma torches sustained by continuous submillimeter radiation in a flow of (a) argon 15 L/min, (b) gas mixture of argon (15 L/min) and air (0.5 L/min), (c) gas mixture of argon (15 L/min) and hydrogen (0.7 L/min), (d) gas mixture of argon (15 L/min) and carbon dioxide (1.5 L/min).*

**Results and discussion**

Measurements of discharge parameters sustained by continuous 263 GHz subterahertz radiation in a hybrid plasmatron were carried out using an argon-air plasma mixture. The argon flow was fixed at 30 L/min and the air volume content varied from 0.7 to 16.4%. The power of the input subterahertz radiation was varied from 680 to 1000 W.

The electron density in the subterahertz plasma torch was measured from the broadening of the Balmer series H$\alpha$ hydrogen line due to the linear Stark effect [33]. For this purpose, a negligible hydrogen amount (<0.05%) was added to the plasma-forming gas mixture at all discharge sustainance modes. Such a small addition of molecular gas did not affect the shape of the discharge and the subterahertz radiation power absorbed by it. The results of spectral measurements presented in Figure 4 correspond to the integral values of electron density averaged over the entire volume of the plasma torch. It can be seen that the discharge sustained in argon flow (with hydrogen added <0.05%) has an integral electron density of $2.5\pm0.5\cdot10^{15}$ cm$^{-3}$, which is 2.8 times higher than the critical electron concentration for the frequency of the electromagnetic wave taking into account the frequency of electron-neutral collisions [34]. Most probably, the electron density has an inhomogeneous spatial distribution, which can be judged indirectly by the presence in the plasma volume of thin filamentary channels elongated along the argon flow direction and perpendicular to the direction of the wave's electric field (Figure 3a). The electron concentration in these filaments can be much higher than the measured integral value. These high local values of electron density may be caused by nonlinear resonance effects of wave electrodynamics arising in a collisionless inhomogeneous argon plasma layer with electron density close to the critical electron density for the frequency of the electromagnetic wave [15-17]. In particular, the formation of strongly supercritical plasma filaments can be associated with the evolution of plasma resonance in a narrow region of singularity of the dielectric permittivity of the plasma [15, 16, 35, 36]. In this case, resonant amplification of the electric field component, which is parallel to the electron density gradient of the inhomogeneous plasma layer, occurs.

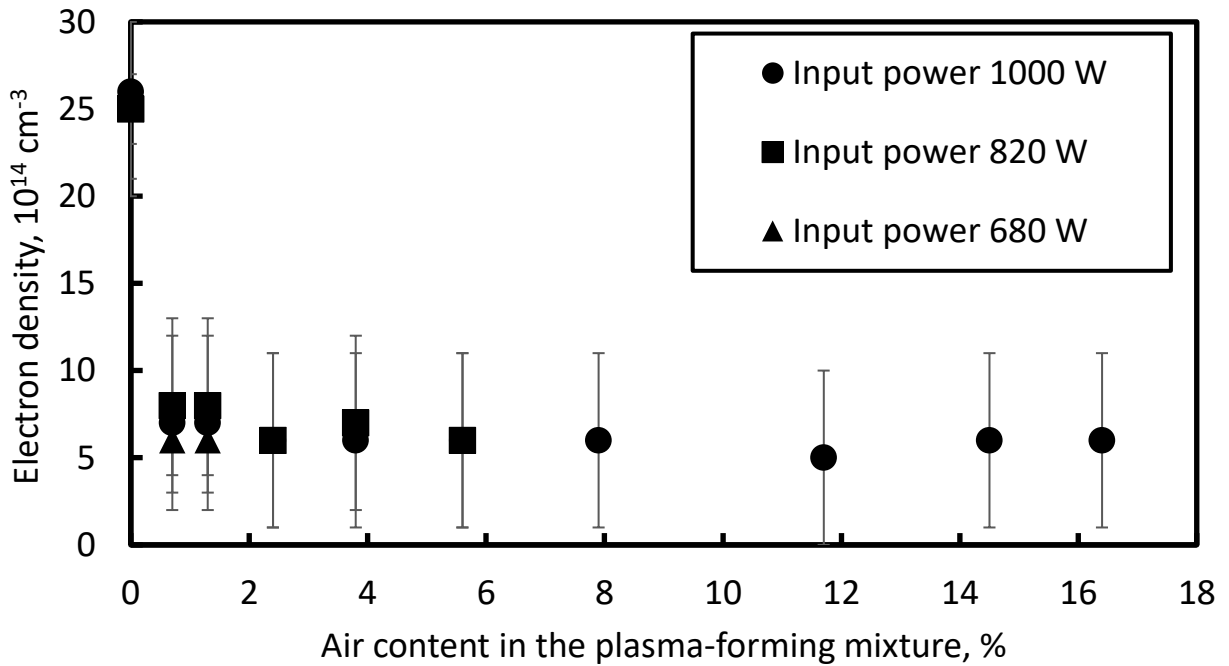

*Figure 4. Integral electron density in the subterahertz discharge as a function of the air content in the argon-air plasma mixture at different heating powers (680, 820, 1000 W).*

The addition of molecular gas to argon can lead to a decrease in the efficiency of the evolution of plasma resonance, since in this case the frequency of electron-neutral collisions increases and the energy threshold of inelastic collisions decreases significantly [34, 37]. Probably, for this reason, it was not possible to visually observe filamentation of discharges sustained in the flow of argon-molecular gas mixture (Figure 3 b, c, d). Therefore, the addition of air to the plasma-forming mixture leads to the decrease in the value of the integral electron density to $7\pm5\cdot10^{14}$ cm$^{-3}$. It can also be seen from Figure 4 that changing the input power of subterahertz radiation and the ratio of components of the plasma-forming mixture did not affect the value of the integral electron density within the measurement accuracy. It can be assumed that, in a subterahertz discharge sustained in the flow of argon and molecular gas, the electron density reaches a near critical value for the frequency of the heating field. A similar result was obtained earlier in a study of the discharge sustained by a focused subterahertz beam in an opposite-directed flow of an argon-carbon dioxide gas mixture at atmospheric pressure [32]. It is worth noting that for the experiments described in that paper, the plasma torch absorbed no more than 15% of the input power, the plasma volume varying from 0.04 to 0.2 $cm^3$. Using a hybrid plasmatron, all other conditions being equal, the volume of the plasma torch is about 6-8 times larger (taking into account the plasma volume filling the electrodynamic "funnel"), which is due to the larger efficiency of absorption of subterahertz radiation by the discharge.

The gas temperature was measured for the argon plasma torch. Emission spectra in the range of 350 - 600 nm with an exposure time of 20 s were obtained using the spectrometer observation channel (Figure 5).

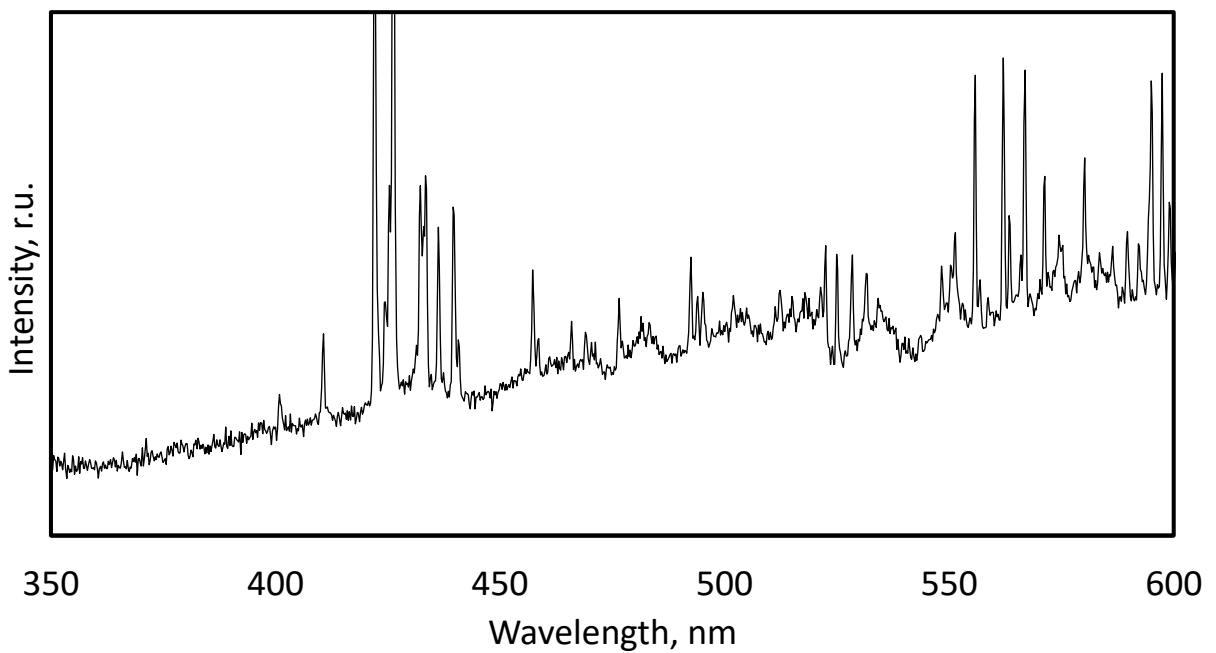

*Figure 5. Characteristic emission spectrum of argon plasma torch.*

In this range there are argon emission lines corresponding to electron transitions of argon atom and a continuous spectrum, which is formed by radiation of heated gas. Argon emission lines were removed from the spectrum using numerical methods and the blackbody continuum was approximated by a function corresponding to Planck's law. The approximation parameters were the greyness factor (a constant multiplier in Planck's law) and temperature. For each separated experimental continuum in the selected spectral range, a definite integral was calculated, which was used to convert the approximated greyness coefficient as a function of temperature. Then, using an iterative procedure, the temperature value of the Planck's function was selected that corresponded to the smallest standard deviation from the experimental continuum.

Figure 6 shows the dependence of the integral gas temperature of the filamentary argon torch on the input power of submillimeter radiation.

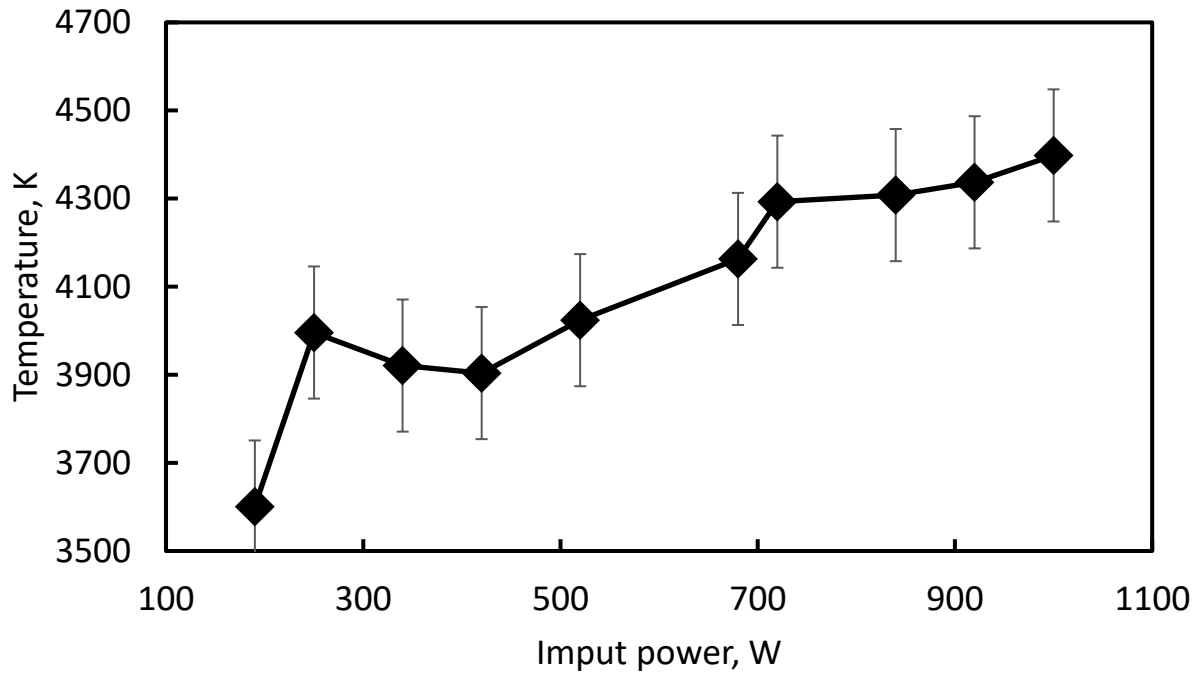

*Figure 6: The dependence of the integral temperature of the argon plasma torch on the input power of submillimeter radiation.*

It can be seen that as the input power increases, the gas temperature slowly increases from 3600 to 4300 K. Therefore, it can be stated that the increase in submillimeter power is mostly spent on increasing the volume of the plasma torch. It is also worth underlining that the obtained values of the gas temperature are an integral average value of the gas temperature in the plasma filaments and the surrounding plasma halo. Based on the literature data on the filamentation mechanisms of microwave discharges [17], we can assume that the gas temperature in the plasma filaments significantly exceeds the gas temperature in the plasma halo.

The power absorbed by the plasma was measured in the regimes described above. Figure 7 shows the results of discharge calorimetry as a function of air content in the plasma mixture at different powers of the input subterahertz radiation. The discharge sustained in the argon flow absorbs about 30% of the input subterahertz power. Adding molecular gas (air) to the plasma-forming mixture leads to an increase in the absorption coefficient up to 62-78% depending on the discharge sustaining mode.

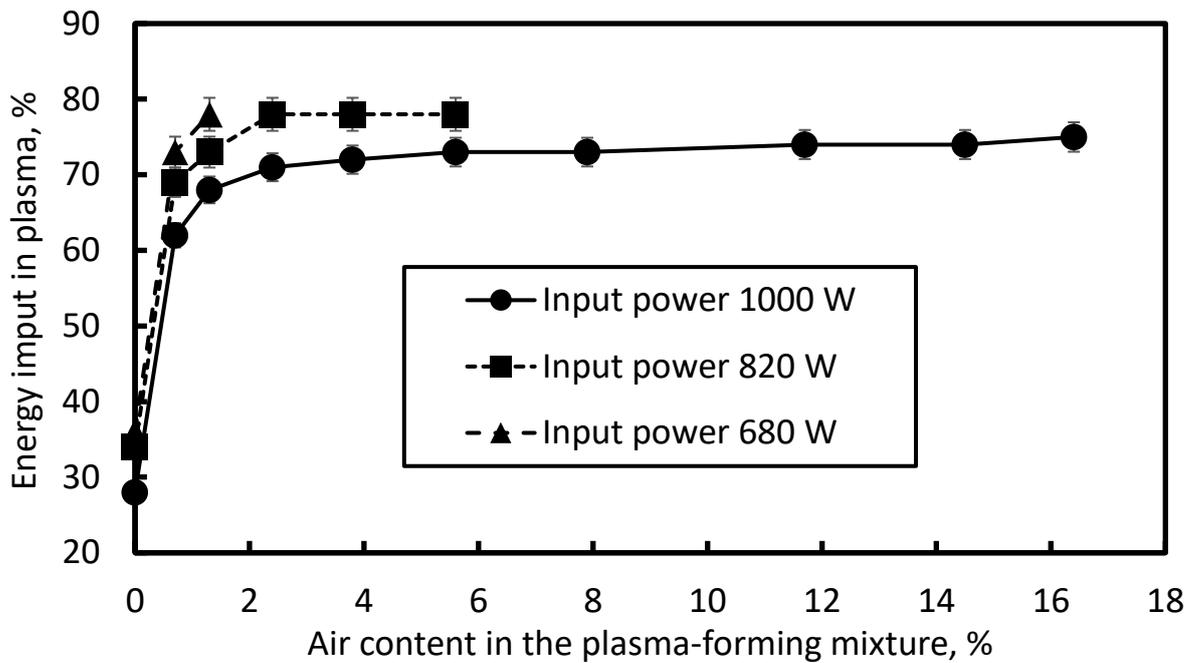

*Figure 7. The dependences of the power part absorbed by the discharge on the air content in the plasma-forming argon-air mixture at different powers of the input subterahertz radiation (680, 820, 1000 W).*

As the air content in the plasma-forming mixture increases, the absorption coefficient reaches saturation, which level is determined by the total power input into the gas-discharge chamber. The difference in the absorption coefficients of submillimeter radiation by the plasma torch sustained in the flow of argon and argon-air gas mixture is apparently due to the reflection of electromagnetic radiation from the plasma layer with supercritical electron density. Indeed, in the experiments with argon plasma torch, about 80% of the submillimeter radiation power not absorbed by the discharge was registered in the upper gas discharge chamber, and 10-20% - in the lower one. Thus, it can be stated that about half of the input electromagnetic radiation power is reflected from the surface of the argon plasma torch. For the discharge sustained in the flow of argon-air gas mixture, almost the opposite configuration was realised: about 20 - 30% of the power not absorbed by the discharge was registered in the upper gas discharge chamber, and 70% - in the lower one. The presence of the molecular component in the plasma-forming mixture affected the discharge shape and electron density distribution so much that the reflection of submillimeter radiation from the torch surface decreased by a factor of 4 - 8 and is about 7 - 12% of the total input power.

For the regimes described above, gas samples were analyzed to determine the conversion of the nitrogen fixation, which characterizes the percentage of nitrogen oxides in the plasma mixture, and the energy efficiency of nitrogen oxide synthesis from atmospheric air, which characterizes the part of energy spent on the synthesis reactions. Component analysis showed that NO is predominantly synthesized in the discharge, the $NO_2$ content is very low and is at the limit of sensitivity of the chromatography-mass spectrometry method. As the submillimeter radiation power increases, the conversion increases from 0.2% to 0.5% (air content in the plasma-forming mixture is 2.4%). Energy efficiency within the accuracy increases from 0.9% to 1.4%. By increasing the air content from 0.7 to 11.7% in the plasma mixture, the conversion of the nitrogen fixation decreases from 0.6% to 0.2%, while the energy efficiency increases from 1% to 1.3%. The obtained results are far from the record values achieved in discharges sustained at atmospheric pressure [38]. Nevertheless, these results are promising because the conversion value varies by a factor of 3 depending on the discharge sustaining parameters, which indirectly indicates a non-equilibrium synthesis mechanism.

**Conclusions**

As the result of this work, a hybrid scheme for organizing a gas discharge sustained by continuous focused submillimeter radiation at 263 GHz has been developed. The combined use of the beam quasi-optical focusing system and the localized electrodynamic structure made it possible to implement effective conditions for sustaining the discharge, which absorbs up to 80% of the input power. It is shown that the discharge sustained in argon flow contains filamentary plasma channels. Formation of this channels can be related to the development of plasma resonance in a collisionless inhomogeneous plasma layer. The principles described in this paper can be a basis for design of subterahertz plasmatrons, which makes a wide range of possibilities for their practical application.

**Acknowledgements**

The authors are grateful to A.Yu. Sozin, O.Y. Chernova, T.G. Sorochkina for chemical analysis of gas samples.

This study was supported by a grant from the Russian Science Foundation (Project No.22-72-00073).

**Declarations**

**Ethical Approval**
not applicable

**Funding**

This study was supported by a grant from the Russian Science Foundation (Project No.22-72-00073).

**Availability**

not applicable